\documentclass{article}

\usepackage{graphicx} 
\usepackage{environ}
\usepackage{listings}
\usepackage[auth-sc]{authblk}
\usepackage{amsmath}
\usepackage[vcentermath]{youngtab}
\usepackage[table]{xcolor}
\usepackage[a4paper]{geometry}
\usepackage[numbers]{natbib}
\usepackage[colorlinks=true,citecolor=red]{hyperref}
\usepackage{relsize}
\usepackage{tikz}
\usepackage{multirow}
\usepackage{multicol}

\usetikzlibrary{shapes,arrows}

\DeclareMathOperator{\sgn}{sgn}
\DeclareMathOperator{\Tr}{Tr}

\NewEnviron{smallequation}{%
    \begin{equation}
    \begin{aligned}
    \scalebox{0.5}{$\BODY$}
    \end{aligned}
    \end{equation}
    }

\newcommand{\adjform}[2][2mm] {
	\vspace{#1}
	\texttt{
		\rowcolors{4}{gray!10}{gray!20}
		\renewcommand{\arraystretch}{1.5}
		\setlength\arrayrulewidth{1px}
		\begin{tabular}{ |c|c|c|c|c|c|c| }
			\hline
			\multicolumn{6}{|c|}{Slot} &  \\
			\cline{1-6}
			\,0\, & \,1\, & \,2\, & \,3\, & \,4\, & \,5\, & \multirow{-2}{*}{Weight}\\
			\hline
			\hline
			#2
		\hline
		\end{tabular}
	}
	\vspace{#1}
}

\newcommand{\twopartdef}[4] {	\left\{\begin{array}{ll}#1 & \mbox{if } #2 \\#3 & \mbox{if } #4 \end{array}\right.}

\title{\hbox{\bfseries Hiding canonicalisation in tensor computer algebra}}
\author{Dominic Price, Kasper Peeters and Marija Zamaklar\\[1ex]\small{\tt
    dominic.t.price@durham.ac.uk}, {\tt kasper.peeters@durham.ac.uk},
  {\tt marija.zamaklar@durham.ac.uk}}
\affil{\small Department of Mathematical Sciences\\Durham University\\South
  Road\\DH1 3LE  Durham\\United Kingdom}
\date{August 25, 2022}

\begin{document}

\maketitle

\begin{abstract}
  Simplification of expressions in computer algebra systems often
  involves a step known as ``canonicalisation'', which reduces
  equivalent expressions to the same form. However, such forms may not
  be natural from the perspective of a pen-and-paper computation, or
  may be unwieldy, or both. This is, for example, the case for
  expressions involving tensor multi-term symmetries. We propose an
  alternative strategy to handle such tensor expressions, which hides
  canonical forms from the user entirely, and present an
  implementation of this idea in the Cadabra computer algebra system.
\end{abstract}

\section{Introduction}

A key part of any symbolic computer algebra system is the ability to
detect equivalence of two mathematical expressions. The common way to
achieve this is to define either a ``canonical form'', such that all
expressions which are equivalent have the same canonical form, or the
weaker concept of ``normal form'', such that all expressions equivalent
to zero are represented by zero (see
e.g.~\cite{geddes1992algorithms}). There is a long history of
canonicalisation algorithms, which of course depend strongly on the
types of expressions one deals with. For any given class of
expressions, it is common to find multiple different canonical or
normal forms in the literature.

For tensor polynomial expressions, which form the topic we deal with
in this paper, there is a fairly extensive literature on
canonicalisation using so-called ``mono-term symmetries''. These are
index permutation symmetries of the expression which take a single
term to a single other term.  A na\"{i}ve method of finding these
symmetries is by considering all the valid permutations of indices,
but most implementations of this are based on the more efficient
Butler-Portugal algorithm~\cite{mans1,mans2}~which is based on the
same concept but scales much better as the number of indices and terms
grow, although still displaying $O(n!)$ behaviour in cases of total symmetry. 
Implementations of this algorithm can be found in Canon~\cite{mans3}~and
xPerm~\cite{DBLP:journals/corr/abs-0803-0862}. An improved version
which uses the same method but with a better $O(n^{1.36})$ complexity
in cases of total symmetry, or large totally symmetric subsets of indices, has
been given in~\cite{Niehoff:2017mbk}.

As mentioned above, these algorithms or their implementations do not
necessarily lead to the same canonical form. This is also true for
``multi-term symmetries'', which map one term to a \emph{sum} of others
(e.g.~Bianchi identities). For canonicalisation subject to such
symmetries, there is far less literature available. Most approaches
treat these multi-term relations as
side-relations~\cite{port2,Martin-Garcia:2007bqa}, and determine
equivalence of expressions subject to these relations using
e.g.~Gr\"obner bases techniques or by defining an explicit canonical
index order for individual tensors~\cite{port2, port3}.  Very few
complete (and maintained) implementations exist.

Many computer algebra systems consider canonicalisation as essentially
the same thing as ``simplification'', that is, the reduction of an
expression to ``the simplest form''. That this is not always helpful
becomes manifest in the case of tensor multi-term symmetries. Canonical
forms will always require that certain simple expressions will be
re-written as more complicated ones with more terms. Many simple
expressions which would appear naturally in a pen-and-paper
computation will turn into much more complicated ones after multi-term
canonicalisation. And even if that does not happen, different people
may not agree on the same definition of ``simplest''.

The present paper builds on the idea that one should let go of the
idea that canonicalisation is the same thing as simplification, and
consider them separately. The question then arises whether
canonicalisation is something which the user should be confronted with
\emph{at all}. We will argue that, in particular for expressions
subject to multi-term symmetries, it is more useful to use
canonicalisation only ``under the hood'', and provide a simplification
algorithm which, instead, aims for a representation of the user's
input which is as compact \emph{and} as close as possible in structure
to the original. We will explain that this idea has the advantage that
canonicalisation is now allowed to produce ugly or very lengthy forms even
for expressions which can be written in a simple way, and one can thus
use algorithms which may not have anything to do with what one would
with pen-and-paper. We discuss such an algorithm for expressions
subject to tensor multi-term symmetries based on Young projectors, and
present a fully-automatic implementation of it in the computer algebra
system Cadabra~\cite{Peeters:2018dyg}.

The outline of this paper is as follows. In
section~\ref{s:review}-\ref{s:young} we summarise the general
formalism behind mono- and multi-term symmetries. We then describe a
generic canonicalisation method based on projection operators in
section~\ref{s:canproj}. The main new contribution is formed by
sections~\ref{sec:hiding_canon} and~\ref{sec:examples}, where we
describe a simplification method based on this underlying
canonicalisation method, and show how it works at the level of
explicit examples. Implementation and complexity details are discussed
in sections~\ref{s:implementation} and~\ref{s:complexity}. We end with
some discussion of extensions of the algorithm in
section~\ref{s:extensions}.

\section{Simplifying generic tensor expressions}
\subsection{Review of tensor canonicalisation methods}
\label{s:review}

The level of support for simplifying expressions based on tensor
symmetries in existing computer algebra software is limited mainly to
the mono-term symmetries of an object, while consideration of the
multi-term symmetries of an object is usually only possible for some
special cases and is generally provided as packages instead of 
core functionality.

For mono-term symmetries, defining a canonical form is simple and
straightforward~\cite{mans1}: given an ordered set of indices, the
canonical representation of an object is that which has the
lexicographically least permutation of its indices which can be
reached using its symmetries. In the case of a tensor with only free
indices this can be mapped to the group theory problem of finding a
representative for the term in the cosets of identical terms; this
idea is extended in \cite{mans2}, where the symmetry of dummy indices
requires the solution to be a double-coset representative. An
alternative is to describe tensors as graphs (avoiding the problem of
dummy index relabelling altogether), as is done e.g.~in
RedBerry~\cite{redberry}.

Defining what a canonical representation means for an object with more
complicated symmetries, such that one term can be equivalent to an
arbitrary number of other terms, is more complicated; proposals have
been made in e.g.~\cite{port2}, although by this definition a single
term may contain multiple terms in its ``canonical'' form. However,
the main problem is that in general the computational expense of
simplifying expressions using multi-term symmetries is very large and
can be impractical even on modern hardware. There are, however, some
approaches which can be taken to solve the problem in specific
situations and there are implementations of these, mainly as packages
for Mathematica and Maple, and mainly for the reduction of Riemann
polynomials.

The method employed by the Invar package~\cite{Martin-Garcia:2007bqa}
(a package that runs on top of xTensor~\cite{DBLP:journals/corr/abs-0803-0862}~or
Canon\cite{mans3}) is to create a rainbow table of possible relations
and, after performing a mono-term canonicalisation, doing a lookup on
the terms in the expression. The scope of this method is obviously
somewhat limiting, typically only working for specific types of tensor
monomials. The major benefit is that this provides a result very
quickly for these common situations.

Another method is to accept a list of symmetries which completely
describe the object and use group algebra techniques to simplify the
expression, as demonstrated by the ATENSOR REDUCE
program\cite{ilyi1}~and the Mathematica package ``Tools of Tensor
Calculus''\cite{e_balf1}. Although this is a completely general method it
suffers from both being far less efficient than the lookup method and
also requiring the user to first calculate a symmetry basis for the
object.

There are other papers describing theoretical approaches to the
problem (such as~\cite{fiedler1997use,fiedler2002ideal}) but without
associated implementations.

\subsection{Mono-term canonicalisation}
\label{sec:monoterm_canon}

Algorithms for tensor canonicalisation typically use the
Butler-Portugal algorithm, which uses the generating set of an
object's mono-term symmetries to find the lexicographically least
permutation of indices~\cite{mans1}. For example, the symmetries of
the Riemann tensor
\begin{equation}
  \begin{aligned}
	R_{a b c d} \equiv -R_{b a c d} \equiv -R_{a b d c}\,,\\[1ex]
	R_{a b c d} \equiv R_{c d a b}\,,
  \end{aligned}
\end{equation}
can be described by the generating set
\begin{equation}
	[ \{-1, (0, 1)\}, \{1, (0, 2)(1, 3)\} ]\,,
\end{equation}
where the first element in each member of the set is the polarity of
the symmetry and the second is the positions of the indices which are
swapped. The relation $R_{a b c d} \equiv -R_{a b d c}$ is obtained by
the application of both these symmetries and therefore does not need
to be a member of this set. From this generating set the full list of
equivalent permutations can be generated; in the case of $R_{c d b a}$
this is
\begin{equation}
	-R_{a b c d},\, R_{a b d c},\, R_{b a c d},\, R_{b a d c},\, -R_{c d a b},\, R_{c d b a},\, R_{d c a b},\, -R_{d c b a}\,,
\end{equation}
from which, assuming no other ordering has been imposed, the least
element is $-R_{a b c d}$ and thus this is calculated as the canonical
representation of $R_{c d b a}$. The 16~remaining index permutations
of the Riemann tensor similarly fall into two lists each containing 8
permutations and there are thus only 3 independent canonical
representations of the Riemann tensor:
\begin{equation}
	\label{eq:riemann_independent}
	R_{a b c d},\, R_{a c d b},\, R_{a d b c}\,.
\end{equation}
By applying this algorithm to all terms in a sum of Riemann tensors it
will therefore be reduced to the simplest possible representation
containing at most 3 terms.

However when considering multi-term symmetries this method is no
longer sufficient for detecting identical terms --- take the simple
example of the first Bianchi identity
\begin{equation}
	\label{eq:first_bianchi}
	R_{a b c d} + R_{a c d b} + R_{a d b c} \equiv 0\,.
\end{equation}
It is clear that application of the Riemann tensor mono-term
symmetries will never yield~\eqref{eq:first_bianchi} as these are the
three independent permutations of the Riemann tensor
from~\eqref{eq:riemann_independent}.  Of course, examples such as this
can be handled as special cases, but in the general case a more robust
solution is required.

\subsection{Tensor symmetries as Young diagrams}
\label{s:young}

The ``canonicalisation'' routine we propose relies on the application
of Young symmetrisers which are informally introduced here. There is
a large number of textbooks on representations of the symmetric group;
much of the following was inspired
by~\cite{birdtracks}~and~\cite{hamermesh}~whilst a terser introduction
to the subject can be found in~\cite{mcnamara}.

The symmetries of objects which are invariant under permutation are
described by the symmetric group $S_n$ whose elements $\sigma_i$
consist of the $n!$ possible permutations of the elements $1, 2,
\dots, n$. These permutations are commonly labelled in cycle notation,
so that the label $i=(145)$ indicates the rearrangement $1 \rightarrow
4$, $4 \rightarrow 5$ and $5 \rightarrow 1$. Some permutations may be
described by disjoint products of cycles such as $(12)(34)$. If a
product of cycles is not disjoint, that is to say more than one cycle
contains the same label, then conventionally the permutations are
performed starting with the rightmost and working left, but it can
also be rewritten as a product of one or more disjoint cycles by
following where each cycle goes to, for example $(123)(24) \equiv
(2413)$.  Permutations can be divided into two categories depending on
whether the number of swaps required to create it is even or odd; the
permutation $(23)$ only requires one swapping ($2 \leftrightarrow 3$)
and is an odd permutation, whilst $(12345)$ is an even permutation
requiring four swaps ($1 \leftrightarrow 2$, $1 \leftrightarrow 3$, $1
\leftrightarrow 4$, $1 \leftrightarrow 5$).  

We can define representations $\psi_i$ of $S_n$ and it is these which
define the symmetry of the object on which the group acts; the
simplest example of a representation of $S_n$ is the \emph{trivial
  representation} which sends each element to the identity operator:
$\psi_{\text{triv}}(\sigma) = \boldsymbol{1}$. Clearly this represents
a completely symmetric object, for instance the metric tensor $g_{\mu
  \nu} \equiv g_{\nu \mu}$ has this representation on $S_2$. Another
very common representation is the \emph{alternating representation}
defined by $\psi_{\text{alt}}(\sigma) = \sgn(\sigma)$ where
\begin{equation}
	\sgn(\sigma) = \twopartdef{1}{\mbox{$\sigma$ is an even permutation\,,}}{-1}{\mbox{$\sigma$ is an odd permutation\,,}}
\end{equation}
so that $\psi_{\text{alt}}(\sigma_{(123)}) = 1$ and
$\psi_{\text{alt}}(\sigma_{(45)}) = -1$.  Objects with this
representation are completely anti-symmetric, like the Levi-Civita
symbol $\epsilon^{a b c} \equiv -\epsilon^{b a c} \equiv \epsilon^{b c
  a} \equiv \dots$ and the structure constants of many Lie
algebras. As these two representations are both one-dimensional,
together they form the complete set of irreps for $n=2$: this is
equivalent to stating that an object with two elements can be
decomposed into a mixture of symmetric and antisymmetric parts, which
is commonly seen in the classic decomposition of a matrix
\begin{equation}
	\label{eq:matrix_decomp}
	M_{a b} \equiv \frac{1}{2}(M_{a b} + M_{b a}) + \frac{1}{2}(M_{a b} - M_{b a})\,.
\end{equation}
Together the two brackets on the right hand side add up to $M_{a b}$,
however, as the first bracket is manifestly symmetric and the second
antisymmetric if the matrix $M_{a b}$ possessed either of these
symmetries one of the brackets would be identically zero.

When $n>2$ these two representations no longer suffice to fully cover
$S_n$, however, the construction of irreps is simplified by the use of
Young tableaux, which are a diagrammatic representation of the irreps
of $S_n$. The construction begins by drawing the \emph{Young diagrams}
for some $n$, and then by filling these diagrams with labels we will
find the complete set of Young tableaux, each of which corresponds to
an irrep. Each diagram $\lambda$ corresponds to a partition of $n$,
commonly denoted $\lambda \vdash n$. In the case of $n=3$ there are
three partitions: $\lambda_1 = (3)$, $\lambda_2 = (2,1)$ and
$\lambda_3 = (1,1,1)$. A Young diagram is a left-justified rows of
cells, where the length of each row is given by one of the members of
the partition and the length of each row is always less than or equal
to the row above. So for $n=3$ the diagrams are
\begin{equation}
	\lambda_1 = \yng(3)\,, ~~
	\lambda_2 = \yng(2,1)\,, ~~
	\lambda_3 = \yng(1,1,1)',.
\end{equation}
The set of all standard tableaux, and thus the irreps of $S_n$, is
given by all the possible standard fillings of these diagrams, which is
done by assigning the labels $1, 2, \dots, n$ to each cell such that
in each row the labels are strictly increasing left to right and in
each column they are strictly increasing top to bottom:
\begin{equation}
	\Lambda_{\text{triv}} = \young(123)\,, ~~
	\Lambda_{\text{std\textsubscript{1}}} = \young(12,3)\,, ~~
	\Lambda_{\text{std\textsubscript{2}}} = \young(13,2)\,, ~~
	\Lambda_{\text{alt}} = \young(1,2,3)\,.
\end{equation}
Each diagram represents a symmetrization along each row and an
antisymmetrization along each column, from which it is easy to see
that $\Lambda_{\text{triv}}$ corresponds to the trivial representation
and $\Lambda_{\text{alt}}$ the alternating representation. The other
two tableaux are known as the \emph{standard representations} and
describe a mixed symmetry that is neither entirely symmetric nor
entirely antisymmetric.

These tableaux can also be used to construct a decomposition, such as
the one in (\ref{eq:matrix_decomp}), by constructing a
\emph{projection operator} for each tableau given by
\begin{equation}
	\label{eq:projector_def}
	P^+(\Lambda)= \frac{1}{N} \prod_{r \in \Lambda}S(r) \prod_{c \in \Lambda}A(c)\,.
\end{equation}
Here $r$ and $c$ are the rows and columns of the tableau, $S$ and $A$
the symmetrization and antisymmetrization operators. $N$ is a
normalisation constant given by the product of the hook-lengths of
each cell, the hook-length of a cell being the number of cells to the
right and below a cell including the cell itself: in the diagram below
each cell contains its hook length
\begin{equation}
	\young(75431,5321,1)\,,
\end{equation}
and the normalisation would be $N = 7\times 5^2 \times 4 \times 3^2
\times 2 \times 1^3 = 12600$. If the order of expanding out the
columns and rows were reversed then we would obtain a different
projector $P^-$; the symmetries become manifest under either
formulation as long as the two types of projectors are not mixed, we
will assume the convention that our projectors are of the form $P^+$.

For the four irreps of $S_3$ we find
\begin{equation}
	\label{eq:n3projections}
	\begin{aligned}
		P\left(~{\young(123)}~\right)T_{a b c} &= \frac{1}{6}(T_{a b c} + T_{a c b} + T_{b a c} + T_{b c a} + T_{c a b} + T_{c b a})\,, \\
		P\left(~{\young(12,3)}~\right)T_{a b c} &= \frac{1}{3}(T_{a b c} + T_{b a c} - T_{c b a} - T_{b c a})\,, \\
		P\left(~{\young(13,2)}~\right)T_{a b c} &= \frac{1}{3}(T_{a b c} + T_{c b a} - T_{b a c} - T_{c a b})\,, \\
		P\left(~{\young(1,2,3)}~\right)T_{a b c} &= \frac{1}{6}(T_{a b c} - T_{a c b} - T_{b a c} + T_{b c a} + T_{c a b} - T_{c b a})\,. \\
	\end{aligned}
\end{equation}
It can be easily calculated that, similarly to (\ref{eq:matrix_decomp}), the sum of these four projectors is $T_{a b c}$ and represents the decomposition of $T_{a b c}$ into symmetric parts.

The symmetry of a general tensor with $n$ indices can be any
combination of the irreps of $S_n$, and in order to determine which
tableaux contribute to a tensor's symmetry this full decomposition can
be written and any terms which do not satisfy the symmetry of the
object discarded. However, while the complete set of Young projectors above
sum up to the identity
\begin{equation}
	\sum_{\Lambda \in \mathrel{\raisebox{0.5pt}{\tiny\yng(1)}}^{\otimes n}} P(\Lambda) = \boldsymbol{1}\,,
\end{equation}
the same is not true of other decompositions of products of Young
tableaux as given by the Littlewood-Richardson rule (a derivation for
which can be found in numerous places); examining the projections
given in (\ref{eq:n3projections}) one can see that
\begin{equation}
	P\left(~{\young(1,2)} \otimes {\young(3)}~\right) 
	\neq 
	P\left(~{\young(13,2)}~\right) + P\left(~{\young(1,2,3)}~\right)\,.
\end{equation}
Another counter-intuitive property of these Young projectors is that they are not in general orthogonal, that is
\begin{equation}
	P(\Lambda_i) P(\Lambda_j) \neq \delta_{i j}\,,
\end{equation}
although this does hold in $\leq 4$ dimensions (see e.g.~\cite{ks-compact}). 

In this case a different construction of the projectors must be
used. There are different constructions for which this and many other
useful properties hold, including the Hermitian KS
construction~\cite{ks-hermitian}, their compact
counterparts~\cite{ks-compact}, and the original orthogonal
construction given by Littlewood~\cite{littlewood-groups}.  It should
be noted that these constructions are far more computationally
intensive than the simple construction described above, and while
properties such as orthogonality and Hermiticity are desirable in
other contexts, they are superfluous to the workings of our algorithm
other than when a symmetry is described by a sum of irreps, in which
case this will be noted.

\subsection{Canonicalisation using projection operators}
\label{s:canproj}

For any tensor symmetries, but in particular for multi-term ones, a
simple way to canonicalise an expression is to project all tensors
using their Young projection operators. The simplicity of this lies in
the fact that it makes all symmetries manifest, and all multi-term
relations will be satisfied identically after the projection. This has
been used in concrete applications in the past (see
e.g.~\cite{Green:2005qr}). This process is however cumbersome to
the user and is often only practical if one has a reasonable idea of what the
answer should be prior to the actual calculation. The huge number of 
terms which can be generated by a Young projection also mean that a
naive approach to the problem can easily lead to space and time constraints
becoming unmanageable.

To illustrate the robustness of this technique, consider the Bianchi
identity from section \ref{sec:monoterm_canon} which the mono-term
algorithms failed to spot.  Fortunately this symmetry of the Riemann
tensor can be made manifest by writing out the Young projection of
each term.  The decomposition of $S_4$ into irreps yields
\begin{equation}
		\young(1234) \oplus \young(012,3) \oplus \young(013,2) \oplus \young(023,1) \oplus 
		\young(02,13) \oplus \young(01,23) \oplus \young(02,1,3) \oplus \young(01,2,3) 
		\oplus \young(03,1,2) \oplus \young(0,1,2,3)\,.
\end{equation} 
Clearly the only one of these terms which satisfies the symmetries of the Riemann tensor is
\begin{equation}
	\Yvcentermath1
	\young(02,13)
\end{equation}
and so the projection of $R_{a b c d}$ consists of the 16 terms 
\begin{equation}\begin{split}
	\frac{1}{12} \Big(  R_{a b c d} - R_{a b d c} - R_{a c d b} + R_{a d c b} - 
	R_{b a c d} + R_{b a d c} + R_{b c d a} - R_{b d c a}  \\[1ex]
	 - R_{c a b d} + R_{c b a d} + R_{c d a b} - R_{c d b a} + R_{d a b c} - 
	R_{d b a c} - R_{d c a b} + R_{d c b a} \Big)\,.
\end{split}\end{equation}
This expression manifestly exhibits all symmetries of the Riemann
tensor.  It is clearly not a very compact canonical form, but it
does allow us to prove the Bianchi identity: by performing the Young
projection of $R_{a c d b}$ and $R_{a d b c}$ as well, all the terms
can be summed together to show that the result is indeed~$0$.

The Young-projected canonical form is thus far from a simplification,
but it can be used to verify the equivalence of two expressions. 
The expression
\begin{equation}
	\label{eq:riemann_two_terms}
	R_{a c d b} + R_{a d b c}
\end{equation}
can be projected and compared against the projection of $R_{a b c d}$
to find that they are identical up to a minus sign, so that the
expression can be substituted for $-R_{a b c d}$ as expected. In the
remainder of this paper we discuss an alternative approach to
simplification, which, similar to the example above, uses the messy
and lengthy Young-projection canonicalisation under the hood to ensure
that any given expressions never uses an over-complete basis of terms.

\subsection{Hiding canonicalisation in simplification}
\label{sec:hiding_canon}

The algorithm we propose based on this mechanism is implemented in
Cadabra~\cite{Peeters:2018dyg}~with the name \texttt{meld}. The
purpose of this algorithm is to provide an all-purpose routine which
can be applied to any expression and ensure that only the minimal set
of basis terms will appear in the output. This general logic applies
to symmetries which go beyond those which can be achieved using Young
projectors, and we will discuss a few of these in
section~\ref{s:extensions}, but the main focus of this paper will be
on tensor expressions.

\begin{figure}[p]
	\centering
	\tikzstyle{decision} = [diamond, draw, fill=lightgray,
	    text width=5em, text centered, node distance=3cm, inner sep=0pt]
	\tikzstyle{block} = [rectangle, draw, fill=white, 
	    text width=8em, text centered, rounded corners, minimum height=4em]
	\tikzstyle{line} = [draw, -latex']
	\tikzstyle{cloud} = [draw, ellipse,fill=darkgray, text=white, node distance=3cm,
	    minimum height=2em]
	\begin{tikzpicture}[node distance = 3cm, auto]
		\node [block] (procinput) {Move terms in input into \texttt{unseen\_terms}};
		\node [cloud, left of=procinput, xshift=-3cm] (input) {Input};
		\node [cloud, right of=procinput] (finished) {Finished};
		\node [decision, below of=procinput] (unseenleft) {Length of \texttt{unseen\_terms} = 0?};
		\node [block, below of=unseenleft] (procA) {Pop first term from \texttt{unseen\_terms} and Young project};
		\node [block, below of=procA] (procB) {Attempt to write as a linear combination of \texttt{unseen\_terms}};
		\node [decision, left of=procB, xshift=-1cm] (solution) {Solution exists?};
		\node [block, above of=solution] (rewrite) {Add contributions from term to scalar prefactors of \texttt{unseen\_terms}};
		\node [block, left of=solution, xshift=-1cm] (add) {Add current Young projection to seen terms};
		
		\path [line] (input) -- (procinput);
		\path [line] (procinput) -- (unseenleft);
		\path [line] (unseenleft) -- node {No} (procA);
		\path [line] (unseenleft) -| node[near start, anchor=north] {Yes} (finished);
		\path [line] (procA) -- (procB);
		\path [line] (procB) -- (solution);
		\path [line] (solution) -- node {Yes} (rewrite);
		\path [line] (solution) -- node {No} (add);
		\path [line] (add) |- (unseenleft);
		\path [line] (rewrite) |- (unseenleft);
	\end{tikzpicture}
	\caption{Visual representation of the logic in the
          \texttt{meld} algorithm.}
	\label{fig:meld_flowchart}
\end{figure}
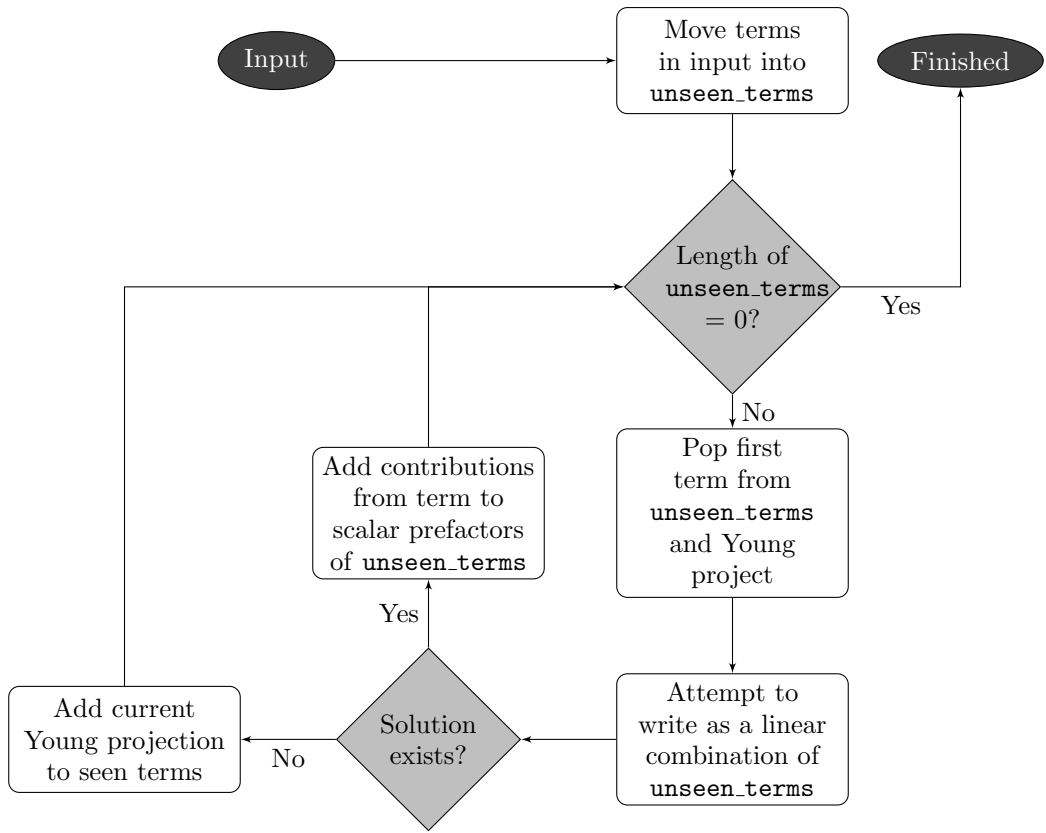

The Young projection routine in \texttt{meld} takes as an input a sum
of terms which are identical in structure modulo index
permutations. The process, visually depicted in
fig.~\ref{fig:meld_flowchart}, is based on updating two stacks:
\verb|seen_terms| and \verb|unseen_terms|. Terms from the input are
added onto the \verb|unseen_terms| stack and one-by-one popped off and
Young projected to see if they can be written as a linear combination
of terms in \verb|seen_terms|. If such a linear combination exists,
then the scalar factors of the terms in \verb|seen_terms| are updated
to include the contribution from the current term, else the current
term is appended to \verb|seen_terms|. The process continues until
there are no more unseen terms.

From this the difference between ``canonicalisation'' and
``simplification'' is made evident: while a traditional
canonicalisation routine will take each term in turn and replace it
with a canonical representation so that at the end all identical terms
can be collected, \texttt{meld} merely detects terms which are
equivalent to each other and collects them together. An example of
where this can result in radically different outputs is found
in~\cite{port2}, where the definition of canonical form causes the
expression
\begin{equation}
	R^{a b c d}R^{e~f}_{~a~c}R_{b f d e}
\end{equation}
to be rewritten in the canonical form (modulo dummy index naming)
\begin{equation}
	R^{a b c d}R^{e~f}_{~a~c}R_{b e d f}	 - \frac{1}{4}R^{a b c d}R^{ef}_{~~ab}R_{c e d f}\,.
\end{equation}
The \texttt{meld} algorithm would instead combine the second expression
into the first.

\subsection{Examples}
\label{sec:examples}

\subsubsection{Mono-term symmetries}

Although specifically designed to act on expressions which can only be 
simplified by the examination of the multi-term symmetries of an object,
\texttt{meld} is also capable of canonicalising expressions where only the 
mono-term symmetries need be considered. It must be noted, however, that whilst
traditional canonicalisation routines, such as the \texttt{canonicalise} function
provided by Cadabra which is based on the xPerm~\cite{DBLP:journals/corr/abs-0803-0862}
canonicalisation algorithm, will return a lexicographically least form of the 
tensor, \texttt{meld} will only ever try to combine existing terms. A simple
example is to look at a fully antisymmetric tensor $A_{a b c}$: the following is
the output of an interactive Cadabra session:

\begin{lstlisting}[numbers=none]
> A_{a b c}::AntiSymmetric;
@ Property AntiSymmetric attached to A_{a b c}. @
> ex := A_{b a c} + A_{c b a};
@ A_{b a c} + A_{c b a} @
> meld(ex);
@ 2A_{b a c} @
> ex := A_{b a c} + A_{c b a};
@ A_{b a c} + A_{c b a} @
> canonicalise(ex);
@ -2A_{a b c} @
\end{lstlisting}

The salient point here is that while both algorithms produce a minimal
representation of the input, as the two algorithms use different
approaches to solving the problem they will often produce a different
output. If the lexicographically canonical form is desired then a call
to the far less computationally expensive \texttt{canonicalise}
algorithm after \texttt{meld} can be used.

A more complex example is found in~\cite[10.5.5]{nakahara}~where it is
shown through algebraic means that
\begin{equation}
	\Tr( F \wedge A \wedge A) \equiv - \Tr(A \wedge F \wedge A)\,,
\end{equation}
where A is a matrix-valued one-form and F is a matrix-valued
two-form. Rewriting this in index notation makes the statement
equivalent to the monomial identity:
\begin{equation}
	F^{a b}{}_{\mu \sigma} A^{b c}{}_{\nu} A^{c a}{}_{\rho} \epsilon^{\mu \sigma \nu \rho} + 
	A^{a b}{}_{\mu} F^{b c}{}_{\nu \sigma} A^{c a}{}_{\rho} \epsilon^{\mu \nu \sigma \rho} \equiv 0\,,
\end{equation}
where $\epsilon^{\mu \nu \sigma \rho}$ is the fully antisymmetric
Levi-Civita symbol. In this formulation it is within the scope of
\texttt{meld} to show:
\begin{lstlisting}[numbers=none]
> \epsilon^{\mu \nu \rho \sigma}::AntiSymmetric;
@ Property AntiSymmetric attached to \epsilon^{\mu \nu \rho \sigma}. @
> ex:= F^{a b}_{\mu \sigma} A^{b c}_{\nu} A^{c a}_{\rho} \epsilon^{\mu \sigma \nu \rho}
      + A^{a b}_{\mu} F^{b c}_{\nu \sigma} A^{c a}_{\rho} \epsilon^{\mu \nu \sigma \rho}:
> sort_product(ex);
@  A^{b c}_{\nu} A^{c a}_{\rho} F^{a b}_{\mu \sigma} \epsilon^{\mu \sigma \nu \rho}
+ A^{a b}_{\mu} A^{c a}_{\rho} F^{b c}_{\nu \sigma} \epsilon^{\mu \nu \sigma \rho} @
> meld(ex);
@ 0 @
\end{lstlisting}
Notice that although \texttt{meld} is able to take advantage of the
fact that the $A^{a b}{}_{\mu}$ tensors are identical to symmetrise in
these objects, \texttt{sort\_product} must first be called to ensure
that the structure of the two monomials matches.

\subsubsection{Bianchi identities}

As described above, the aspect of \texttt{meld} which makes it far
more general than \texttt{canonicalise} is its ability to simplify
expressions for which only considering the mono-term symmetries is not
enough. An example shows the difference:
\begin{lstlisting}[numbers=none]
> R_{a b c d}::RiemannTensor.
@ Property TableauSymmetry attached to R_{a b c d}. @
> ex := R_{a b c d} + R_{a c d b} + R_{a d b c};
@ R_{a b c d} + R_{a c d b} + R_{a d b c} @
> canonicalise(ex);
@ R_{a b c d}-R_{a c b d} + R_{a d b c} @
> meld(ex);
@ 0 @
\end{lstlisting}
Here the \texttt{canonicalise} routine is only able to consider each
term individually, making the most canonical form it can find for
each, whilst \texttt{meld} recognises the Bianchi identity, although
of course \texttt{meld} has no knowledge of specific identities but
relies of the Young projection to make all such identities
manifest. It can also take advantage of this if there are arbitrary
scalar factors attached to each term
\begin{lstlisting}[numbers=none]
> ex := a_1 * R_{a b c d} + a_2 * R_{a c d b} + a_3 * R_{a d b c};
@ a_{1} R_{a b c d} + a_{2} R_{a c d b} + a_{3} R_{a d b c} @
> meld(ex);
@ (a_{1}-a_{3}) R_{a b c d} + (a_{2}-a_{3}) R_{a c d b} @
\end{lstlisting}
In this example the expression has been reduced to the fewest possible
number of terms required to represent it making use of the Bianchi
identity.

\subsubsection{Tensor polynomials}

The real convenience of \texttt{meld} comes when considering
polynomials of objects which admit multi-term symmetries as the number
of terms in their Young projection increases exponentially in the
order of the polynomial, however this complexity can be hidden from
the user. A good example of this is the following identity involving
four Riemann tensors~\cite{Peeters:2000qj,Frolov:2001jh}, 
\begin{multline}
		R_{p q r s} R_{p t r u} R_{t v q w} R_{u v s w} - R_{p q r s} R_{p q t u} R_{r v t w} R_{s v u w} \\
	          - R_{m n a b} R_{n p b c} R_{m s c d} R_{s p d a} + \frac{1}{4} R_{m n a b} R_{p s b a} R_{m p c d} R_{n s d c} \equiv 0\,.
\end{multline}
Here there are 65,556 terms in the Young projection of each monomial,
but this complexity is hidden in the implementation of the algorithm
\begin{lstlisting}[numbers=none]
> R_{m n p q}::RiemannTensor;
@ Property RiemannTensor attached to R_{m n p q}. @
> ex:= R_{p q r s} R_{p t r u} R_{t v q w} R_{u v s w}
|      - R_{p q r s} R_{p q t u} R_{r v t w} R_{s v u w}
|      - R_{m n a b} R_{n p b c} R_{m s c d} R_{s p d a}
|      + (1/4) R_{m n a b} R_{p s b a} R_{m p c d} R_{n s d c}:
> meld(ex);
@ 0 @
\end{lstlisting}
It would also be possible to prove this identity by expanding each
monomial separately and then using a traditional canonicalisation
algorithm to simplify each bracket before distributing the result,
however these extra steps are hidden by using \texttt{meld}. The
simplification (to zero) is thus manifestly different from the
internal (but hidden) canonicalisation process.

As mentioned above, \texttt{meld} attempts to combine terms together
in the way in which they appear in the input, and therefore the order
in which terms in a sum are placed can change the output. This is
clearly visible in the two different lines below,
\begin{equation}
  \begin{aligned}
    R_{i j k l} R_{i j k l} + R_{i j k l} R_{i k j l} & \rightarrow  \frac{3}{2} R_{i j k l} R_{i j k l}\,, \\
    R_{i j k l} R_{i k j l} + R_{i j k l} R_{i j k l} & \rightarrow 3 R_{i j k l} R_{i k j l} \,.
  \end{aligned}
\end{equation}
In both cases, \texttt{canonicalise} is unable to detect the relationship
\begin{equation}
	R_{i j k l}R_{i j k l} \equiv 2 R_{i j k l} R_{i k j l}
\end{equation}
which, although initially appearing to be a mono-term relation,
contains a hidden Bianchi identity.

Of course the logic of \texttt{meld} works equally well for
expressions which contain multiple types of fields. The following
example is taken from~\cite{Liu:2019ses}, which deals with monomials
of the symbolic form $H^2 R^3$, where $H$ is a completely
anti-symmetric field strength and $R$ is the Riemann tensor. Table~3
of~\cite{Liu:2019ses} shows an expansion of certain contractions in
terms of a basis, given explicitly in the appendix of that paper. With
\texttt{meld} we can construct a basis explicitly along the following
lines:
\begin{lstlisting}[numbers=none]
> H_{a b c}::AntiSymmetric;
> t8t8H2R3:= t8_{m1 m2 m3 m4 m5 m6 m7 m8} t8_{n1 n2 n3 n4 n5 n6 n7 n8}
             H_{m1 m2 a} H_{n1 n2 a}
             R_{m3 m4 n3 n4} R_{m5 m6 n5 n6} R_{m7 m8 n7 n8};
> t8rule:= t8_{m1 n1 m2 n2 m3 n3 m4 n4} =
           -2 \delta_{n2 m1} \delta_{n1 m2} \delta_{n4 m3} \delta_{n3 m4}
           + ...
> substitute(t8t8H2R3, t8rule)
> distribute(_)           
> eliminate_kronecker(_)  
> meld(_)
@ 12 H_{m_4 m_3 a} H_{n_4 n_3 a}
     R_{m_3 m_4 n_3 n_4} R_{m_8 m_7 n_8 n_7} R_{m_7 m_8 n_7 n_8} + ... @
\end{lstlisting}
This produces a minimal number of terms (six), which does not, however,
use the same basis terms as in~\cite{Liu:2019ses}. If one wanted to
use the particular basis of~\cite{Liu:2019ses}, one could instead
write down a generic linear combination of those basis terms, subtract from
that the expression above, and run \texttt{meld} on the sum.
\begin{lstlisting}[numbers=none]
> expansion := c_1 R_{m n p s} R_{a b c d} R_{a b c d}
               + c_2 R_{m n p a} R_{s b c d} R_{a b c d} + ...;
> zero = expansion - t8t8H2R3;
> meld(zero);
@ (c_1 + 1/8) H_{m n q} H_{p s q} R_{a b c d} R_{a b c d} R_{m n p s} + ...
\end{lstlisting}
The condition that this vanishes then fixes the coefficients~$c_i$ in
agreement with the result of~\cite{Liu:2019ses}. In this way
\texttt{meld} can be used both to generate a basis and to verify an
expansion in terms of an already determined basis.

\subsubsection{Other complex tableaux}

Whilst the most common example of an object with a multi-term is the
Riemann tensor, there are of course many other possible tableaux
shapes admitting a variety of identities. One example is the tensor
$T_{a b c d}$ with the associated tableau
\begin{equation}
	\Yvcentermath1
	\young(013,2)
\end{equation}
from which using \texttt{meld} we can confirm also satisfies the
Bianchi identity
\begin{equation}
	T_{a b c d} + T_{a c d b} + T_{a d b c} = 0\,.
\end{equation}
Similarly other tableau shapes such as $T_{a b c d e}$ described by
\begin{equation}
	\Yvcentermath1
	\young(130,24)\,,
\end{equation}
which is beyond the scope of most modern computer algebra systems to
fully canonicalise can be shown to satisfy the relation
\begin{equation}
	\label{eq:double_bianchi}
	T_{a b c d e} + T_{a b d e c} + T_{b a e c d} + T_{c a d b e} = 0\,.
\end{equation}
Such ``hook tableaux'' are useful as they describe the symmetry of the
covariant derivative of the Riemann tensor with a metric connection:
this can be seen by computing the Littlewood-Richardson decomposition
of the direct product of the two tableaux associated with~$\nabla_{e}$
and~$R_{a b c d}$:
\begin{equation}
	\young(e) \otimes \young(ac,bd) = \young(ace,bd) \oplus \young(ac,bd,e)\,.
\end{equation}
The second Bianchi identity
\begin{equation}
	\nabla_e R_{a b c d} + \nabla_c R_{a b d c} + \nabla_d R_{a b e c}\,,
\end{equation}
is only satisfied by the first term in the decomposition and thus is
the only term which describes the symmetry of $\nabla_e R_{a b c
  d}$. Repeating this process~\cite{fulling, fiedler2003}~reveals that
the $n$th covariant derivative $\nabla_{e_1} \nabla_{e_2} \dots
\nabla_{e_n} R_{a b c d}$ is described by
\begin{equation}
	\overbrace{
		\yng(4,2) \mathrel{\raisebox{7pt}{\dots} \raisebox{3pt}{\yng(1)}}
	}^{\text{$n+2$ boxes}}
\end{equation}

In fact \texttt{meld} can take advantage of this fact to canonicalise
expressions involving covariant derivatives of Riemann tensors, which
we illustrate with an example from~\cite{invar2}:
\begin{equation}
	R^{abcd;e}_{~~~~~~\,a} R_{be}^{~~fg;hi} R_{cfgi;dh} = \frac{1}{8} R^{abcd;e}_{~~~~~~\,e} R_{ab}^{~~fg;hi} R_{cdfg;ih}\,,
\end{equation}
which can be shown in Cadabra with
\begin{lstlisting}[numbers=none]
> D{#}::Derivative.
> R_{a b c d}::RiemannTensor.
> ex := D_{e}{D_{a}{R_{a b c d}}} D_{h}{D_{i}{R_{b e f g}}} D_{d}{D_{h}{R_{c f g i}}} - 
|	1/8 D_{e}{D_{e}{R_{a b c d}}} D_{h}{D_{i}{R_{a b f g}}} D_{i}{D_{h}{R_{c d f g}}};
> meld(ex);
@ 0 @
\end{lstlisting}

\subsection{Implementation details}
\label{s:implementation}

While a high-level description of how the algorithm works has been
presented in section~\ref{sec:hiding_canon}, in this section some of
the lower-level implementation details of the \texttt{meld} algorithm
are given.

\subsubsection{Data storage}

In Cadabra, expressions are implemented as \texttt{Ex} objects which
are trees consisting of nodes which contain a name, rational
multiplier and information such as bracket type and parent relation
which are only set for some node types. There are three main drawbacks
to this storage format when considering storing Young
projections. Most evidently, a collection of terms belonging to a
Young projection will share the exact same structure except for index
permutations. Thus the majority of the stored information is
superfluous. Moreover, nodes are stored in a linked list which causes
sibling traversal, such as iterating over the indices of an object, to
be very likely to cause cache misses. Finally, dummy index naming is
not irrelevant as $T_{a a}$ and $T_{b b}$ have different
representations in this format even though they are identical.

In order to solve these problems a different storage format is
introduced for the internals of the \texttt{meld} algorithm. This format
stores the \texttt{Ex} object of a tensor monomial once per projection,
alongside a list of \emph{adjacency} representations of the index
permutations in the projection, which is a mapping of the actual
indices to a list of integers.  To illustrate this with an example,
consider the expression $R_{a b c d}F_{c d}$ which contains two free
indices $a$ and $b$ as well as two contractions. Internally, before
projecting the term \texttt{meld} stores the \texttt{Ex} object
representing this along with a single adjacency list

\adjform{ -1 & -2 & 4 & 5 & 2 & 3 & 1 \\ }

The slots represent zero-based pointers into the index structure of
the expression, so for the above expression slots 0-3 point to the
four indices of $R_{a b c d}$ and slots 4-5 the two indices of $F_{c
  d}$. The negative integers represent unique free indices and the
positive integers represent dummy indices by acting like pointers and
indicating that the index in that slot is contracted with the index at
the slot it ``points'' to: the value of 4 in slot 2 indicates that
slots 2 and 4 are contracted representing the two $c$ dummy indices in
the expression. For consistency slot 4 contains the value 2 completing
the contraction.

Each adjacency list is also assigned a weight, initially set to 1,
which keeps track of the relative contributions each term has towards
the overall object.

\subsubsection{Symmetrization}

Symmetrization is performed by decomposing the associated tableaux
into columns and rows and symmetrizing along these. Assuming the
tensor $R_{a b c d}$ has been defined with the tableau
\begin{equation}
	\young(02,13)\,,
\end{equation}
the algorithm successively applies the two antisymmetrisers $[0,1]$
and $[2,3]$ and the two symmetrisers $(0,2)$ and $(1,3)$ to all the
adjacency lists:
\begin{multicols}{2}
		\noindent
		\begin{minipage}[t][.22\textheight][t]{\columnwidth}
		\vspace{-\topskip}
			(1) Antisymmetrise in (0,1)
		
			\adjform{ 
				-1 & -2 & 4 & 5 & 2 & 3 & 1 \\
				-2 & -1 & 4 & 5 & 2 & 3 & -1 \\
			}
		\end{minipage}
	
		\columnbreak
	
		\noindent
		\begin{minipage}[t][.22\textheight][t]{\columnwidth}
		\vspace{-\topskip}
			(2) Antisymmetrise in (2,3)
	
			\adjform{ 
				-1 & -2 & 4 & 5 & 2 & 3 & 1 \\
				-2 & -1 & 4 & 5 & 2 & 3 & -1 \\
				-1 & -2 & 5 & 4 & 3 & 2 & -1 \\
				-2 & -1 & 5 & 4 & 3 & 2 & 1 \\
			}
		\end{minipage}
\end{multicols}

\begin{multicols}{2}
		\noindent
		\begin{minipage}[t][.33\textheight][t]{\columnwidth}
		\vspace{-\topskip}
			(3) Symmetrise in (0,2)
		
			\adjform{ 
				-1 & -2 & 4 & 5 & 2 & 3 & 1 \\
				-2 & -1 & 4 & 5 & 2 & 3 & -1 \\
				-1 & -2 & 5 & 4 & 3 & 2 & -1 \\
				-2 & -1 & 5 & 4 & 3 & 2 & 1 \\
				4 & -2 & -1 & 5 & 0 & 3 & 1 \\
				4 & -1 & -2 & 5 & 0 & 3 & -1 \\
				5 & -2 & -1 & 4 & 3 & 0 & -1 \\
				5 & -1 & -2 & 4 & 3 & 0 & 1 \\
			}
		\end{minipage}
	
		\columnbreak
	
		\noindent
		\begin{minipage}[t][.33\textheight][t]{\columnwidth}
		\vspace{-\topskip}
			(4) Symmetrise in (1,3)
	
			\adjform{ 
				-1 & -2 & 4 & 5 & 2 & 3 & 1 \\
				-2 & -1 & 4 & 5 & 2 & 3 & -1 \\
				-1 & -2 & 5 & 4 & 3 & 2 & -1 \\
				\multicolumn{7}{|c|}{\emph{8 terms omitted...}} \\
				4 & 5 & -1 & -2 & 0 & 1 & 1 \\
				4 & 5 & -2 & -1 & 0 & 1 & -1 \\
				5 & 4 & -1 & -2 & 1 & 0 & -1 \\
				5 & 4 & -2 & -1 & 1 & 0 & 1 \\
			}
		\end{minipage}
\end{multicols}
\noindent yielding the 16 terms as expected. Notice that the
projectors are not normalised; this is in order to allow the weights
to be combined using integer arithmetic instead of the more expensive
rational arithmetic which would be required, and makes no difference
to the result as all terms accumulate the same normalisation and so it
only produces a global constant which can be factored out.

The \texttt{meld} algorithm can also accept objects which are defined
with products or sums of tableaux. If products of tableaux, such as
\begin{equation}
	\young(0,1) \otimes \young(2,3)
\end{equation}
are provided, then the symmetrisers and antisymmetrisers are
constructed and again applied in turn.

If a sum of tableaux is provided then the construction of Hermitian
projectors given in~\cite{ks-hermitian}~is used to compute the
symmetriser of each term, for example
\begin{equation}
	\begin{aligned}
	H\left(~{\young(02,1)} \oplus {\young(0,1,2)}~\right) 
		&= 	H\left(~{\young(02,1)}~\right) + H\left(~{\young(0,1,2)}~\right)  \\
		&= 	P\left(~{\young(0,1)}~\right) P\left(~{\young(02,1)}~\right) P\left(~{\young(0,1)}~\right) + 
			P\left(~{\young(0,1)}~\right) P\left(~{\young(0,1,2)}~\right) P\left(~{\young(0,1)}~\right)
	\end{aligned}
\end{equation}
where $H$ is a Hermitian projector and $P$ the construction given in
(\ref{eq:projector_def}). In this instance the normalisation is
relevant to ensure the correct mixing of terms, however it is still
desirable to use integer arithmetic for efficiency and so each term is
multiplied by the product of the all the normalisation constants so
that
\begin{equation}
	\frac{1}{N_1}P_1 + \frac{1}{N_2}P_2 + \dots + \frac{1}{N_k}P_k
\end{equation}
becomes
\begin{equation}
	(N_2 N_3 \dots N_k) P_1 + (N_1 N_3 \dots N_k) P_2 + \dots + (N_1 N_2 \dots N_{k-1}) P_k
\end{equation}
where the brackets are guaranteed to be integer as each $N_i$ is a
product of integer hook-lengths.

\subsubsection{Detection of linear dependence}

The algorithm looks as each term in turn, calculating its Young
projection as above and checking if it is a linear combination of
previously encountered terms by solving the equation
\begin{equation}
\label{eq:linterms}
	L_1 (c_{11} T_1 + \ldots + c_{n1} T_n) + L_2 (c_{12} T_1 + \ldots + c_{n2} T_n) + \ldots = \lambda_1 T_1 + \ldots + \lambda_n T_n\,.
\end{equation}
In this equation each bracket is a previously calculated projection of
an input expression $T_i$ which produces a sum of all the possible
index permutations $T_1, T_2, \dots, T_n$ with coefficients $c_{ij}$
which may be zero, and the variables $L_i$ are to be determined. The
right hand side is the Young projection of the current term which we
wish to express as a linear combination of the other projections.

As each $T_i$ is linearly independent, by comparing coefficients this
equation can be rearranged into the matrix equation $C\vec{L} =
\vec{\lambda}$:
\begin{equation}
	\label{eq:matrixterms}
	\begin{bmatrix}
	c_{11} & c_{12} & \dots & c_{1k} \\
	c_{21} & \ddots & & \\
	\vdots & & & \\
	c_{n1} & & & c_{nk}
	\end{bmatrix}
	\begin{bmatrix}
	L_1  \\
	L_2  \\
	\vdots \\
	L_k
	\end{bmatrix}
	=
	\begin{bmatrix}
	\lambda_1  \\
	\lambda_2  \\
	\vdots \\
	\lambda_n
	\end{bmatrix}\,.
\end{equation}
The matrix $C$ is $n \times k$ and so the solution set depends on the
relation between these two quantities. The columns of the matrix are
always linearly independent as the vector $\vec{\lambda}$ is only
appended to the matrix if it cannot be expressed as a linear
combination of its columns. For $n=k$ there is therefore a unique
solution and so $k$ (the number of terms projected so far) can never
get larger than $n$ (the total number of possible index
permutations). The initial matrix is constructed when $k=1$, and as
all projections contain at least one term we therefore conclude $k
\leq n$.

In fact, as $n$ grows factorially with the number of indices it is
normally much larger than $k$ and the system is very
over-constrained which makes solving the system computationally
expensive. In order to mitigate this problem, the approach we
take is to truncate the columns of $C$ so that it only contains $k$
rows ensuring that linear independence between the columns is
preserved. This is then solved to produce a potential solution
$\vec{L}'$. To determine if it is an actual solution to
(\ref{eq:matrixterms}), $\vec{L}'$ is then substituted back into
(\ref{eq:linterms}) to ensure that it is a real solution; if not then
the algorithm discards the solution $\vec{L}'$ and concludes that
$\vec{\lambda}$ should be appended to $C$.

\subsubsection{Optimisations}
\label{s:optimisations}

Even though the data structure we use reduces the memory footprint of
the projections, their calculation is still an expensive process and
various ways of minimising the complexity of this process have been
considered. The current implementation focuses on two categories of
optimisations: calculating independent symmetrisers and cancellation
of similar symmetrisers. These are perhaps best illustrated by
example. Consider
\begin{equation}
	R^{\mu}_{~\nu \rho \lambda} F_{\mu}^{~\nu} \varepsilon^{i j l k}\,,
\end{equation}
where $R^{\mu}_{~\nu \rho \lambda}$ has the Riemann tensor symmetries
and both $F_{\mu}^{\nu}$ and $\varepsilon^{i j k l}$ are totally
antisymmetric. The projector in terms of index slots is represented by
the product of symmetrisers
\begin{equation}
		\overbrace{
			\young(0,1) \otimes \young(2,3) \otimes \young(02) \otimes \young(13)
		}^{^{=~\young(02,13)}} \otimes \young(4,5) \otimes \young(6,7,8,9)\,.
\end{equation}
We first notice that two symmetrisers commute if they share no
indices. This means that we can bring the antisymmetriser
${\footnotesize\young(45)}^{\,T}$ to the front of the
expression. However, the two slots it contains are contracted with the
slots in the symmetriser $\footnotesize{\young(01)}^{\,T}$ and
therefore the application of one followed by the other is identical to
only applying one allowing us to remove one of these symmetrisers from
the expression. Note that this is only possible because we were able
to pull both symmetrisers to the front of the expression allowing us
to apply them both first, as if some other symmetriser were applied
before these then the structure of the dummy indices would be broken
and we could no longer rely on the indices in slots $(4,5)$ being
contracted with slots $(0,1)$.

Further to the above, we can also see that the symmetriser
$\footnotesize{\young(6789)}^{\,T}$ shares no slots with any other
symmetriser and can be freely commuted through the expression, in
addition to which its indices are not contracted with any other
indices in the expression. It therefore has no interaction with any
other symmetrisers and will add nothing more than an overall factor of
$\pm 1$ to the term, so instead of projecting it and increasing the
total number of terms by $4!$ we simply sort the indices in these
positions, multiply the weight of the initial term by $-1$ if this
sorting corresponds to an odd permutation, and then drop the
symmetriser from the expression. This reduces the total projector to
\begin{equation}
	\young(0,1) \otimes \young(2,3) \otimes \young(02) \otimes \young(13)\,,
\end{equation}
i.e.~only the symmetry of the Riemann-like tensor needs to be calculated.

Other optimisations are possible, for instance the cancellation rules
given in~\cite{ks-compact}, which apply in particular to the
combinations of symmetrisers which result from the construction of a
Hermitian projector.

\subsection{Complexity}
\label{s:complexity}

In the literature on tensor canonicalisation, a lot of attention has
been paid to the complexity of various algorithms. While that is
certainly a worthwhile mathematical aim, the number of practical
research problems in which e.g.~Riemann tensor monomials of order 10
or higher appear, is limited. Our approach has been that for many
situations, it is more useful to have a slow but completely
algorithmic solution than a fast solution which only covers a limited
number of cases or needs tuning by hand. Nevertheless, we should make
some comments about the complexity of our algorithm to put it in context.

\begin{figure}[t]
\includegraphics[width=0.7\textwidth]{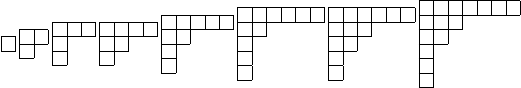}
\centering
\caption{Examples of tableaux which minimise the number of terms in
  the Young projection for $n$ indices.}
\label{fig:mintabs}
\end{figure}

The complexity of the algorithm is dependent on two main factors, the number of terms in
the expression being acted on and the shape of the tableau associated with each term. The shape 
of a tableau $Y$ can be described by a list $(r_1, r_2, \dots, r_n)$ where
each $r_k$ is the number of boxes in the $i$th row of the tableau and thus represents
and object with $n = \sum r_k$ indices. Similarly $Y$ can be described by its columns
$(c_1, c_2, \dots, c_{r_1})$ where each $c_k$ can be computed
by the number of rows which are at least $k$ cells long
\begin{equation}
	c_k = \sum_{i=1}^{n}\Theta(r_i - k), ~~~~ \Theta(x)=\twopartdef{1}{x \geq 0}{1}{\mbox{otherwise}}\,.
\end{equation}
The total number of terms generated by the Young projection operator
of $Y$ is given by the product of the numbers of terms generated by
the (anti)symmetrization of each row and column
\begin{equation}
	N(Y) = \prod_{k=1}^{n}(r_k!)\prod_{k=1}^{r_1}(c_k!)
\end{equation}
although this does not necessarily equal the total number of terms in
the projection, for example in the case of a totally symmetric tensor
$S_{a b a b}$, $4!$ operations are required to produce the resulting 2
terms $\frac{1}{2}(S_{a b a b} + S_{a b b a})$. Clearly the shape of
the tableau plays a crucial role; some examples with a minimal number
of terms in the projection are shown in fig.~\ref{fig:mintabs}.

\begin{figure}[p]
	\includegraphics[width=0.7\textwidth]{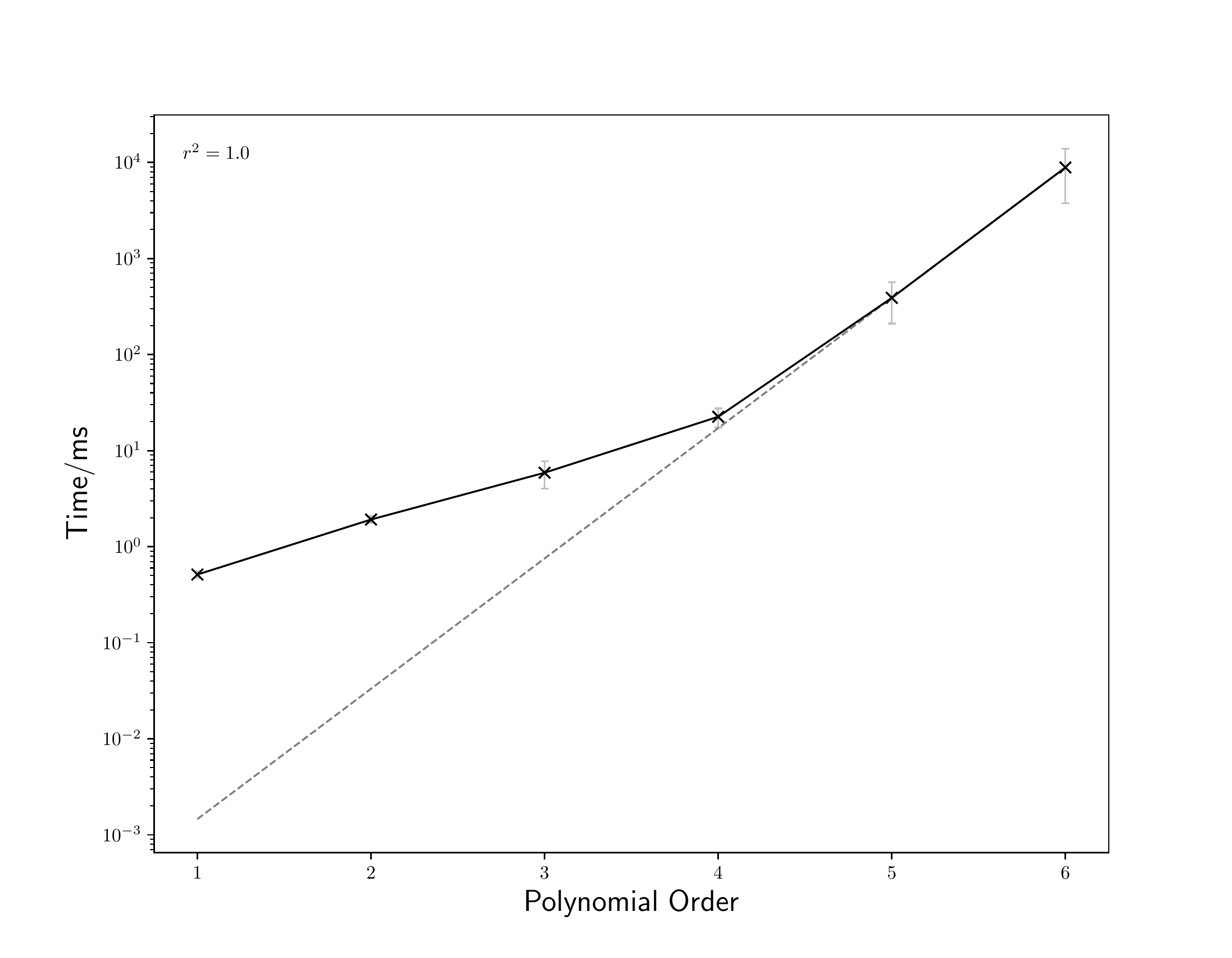}
	\centering
	\caption{Run-time performance of \texttt{meld} as the number
          of terms in a tensor polynomial increases.}
	\label{graph:polyorder}
        \vspace{1cm}
        
	\includegraphics[width=0.7\textwidth]{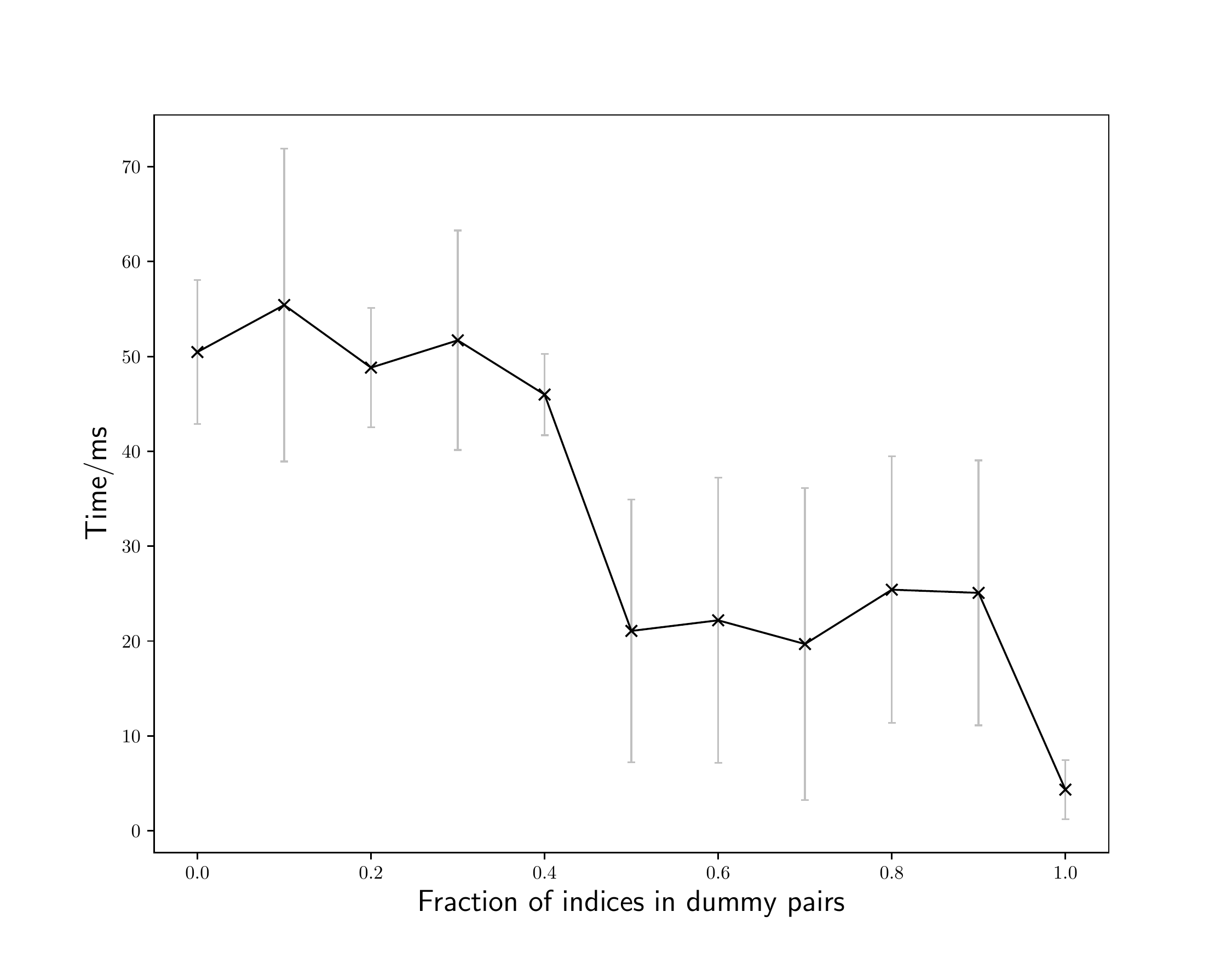}
	\centering
	\caption{Run-time performance of \texttt{meld} as the number
          of dummy indices relative to free indices increases. Because
          of the fact that dummy index relabelling leaves the stored
          expression invariant, performance generically improves when
          a larger fraction of the indices is dummy.}
	\label{graph:dummy}
\end{figure}

\begin{figure}[p]
	\includegraphics[width=0.7\textwidth]{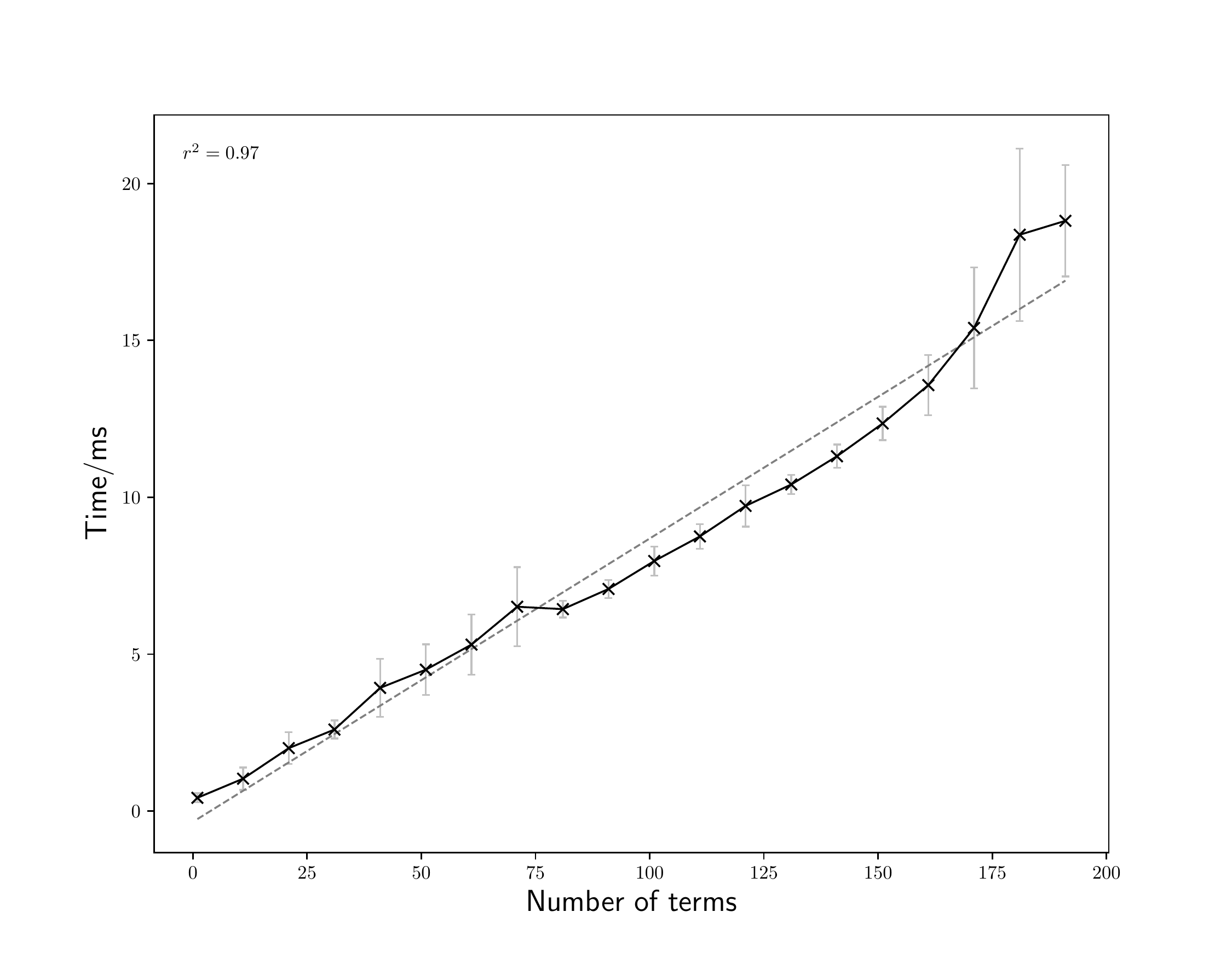}
	\centering
	\caption{Run-time performance of \texttt{meld} as number of terms increases showing the linear relation.}
	\label{graph:n_terms}
        \vspace{1cm}

        \includegraphics[width=0.7\textwidth]{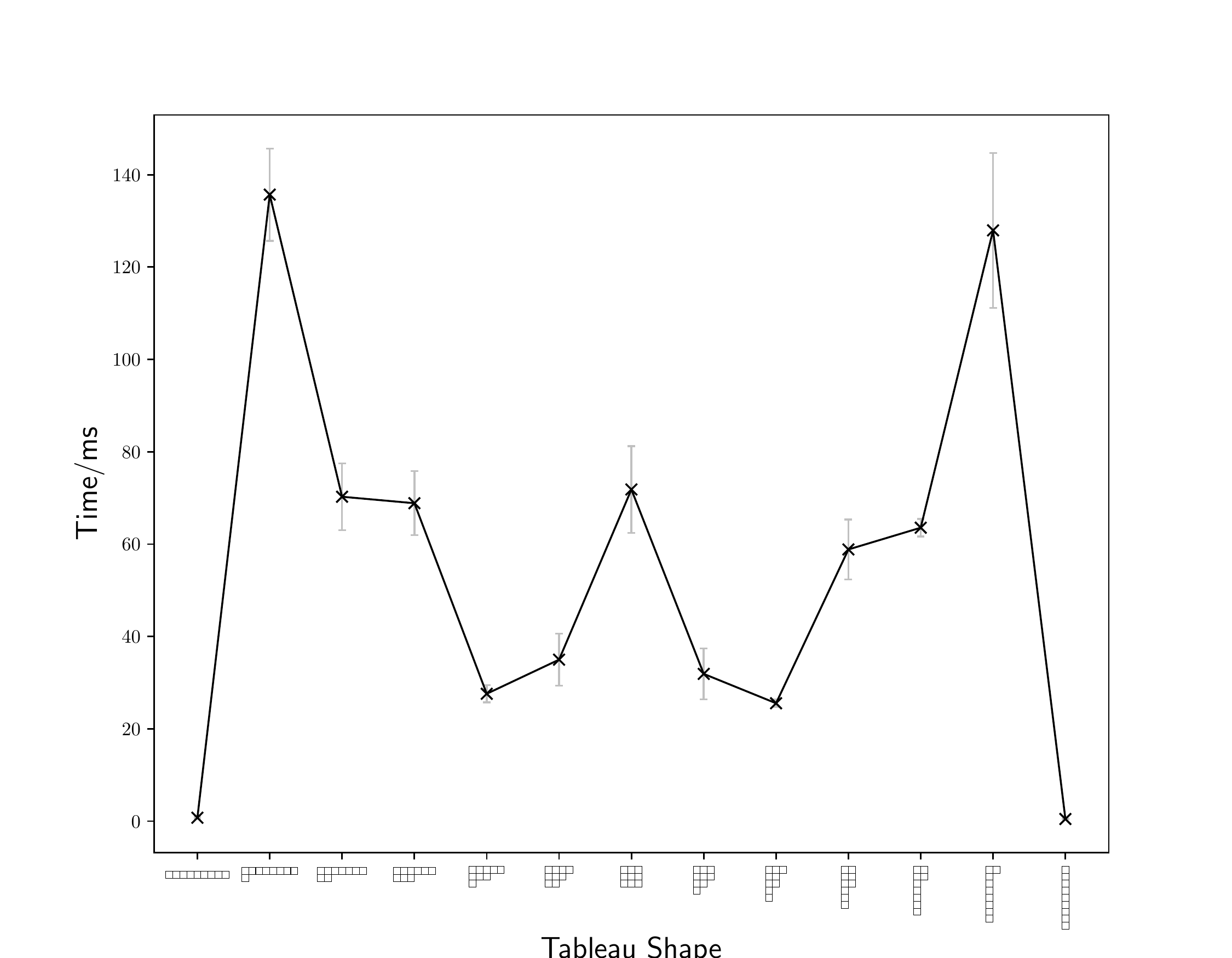}
	\centering
	\caption{Run-time performance of \texttt{meld} as the shape of a tableau with 9 cells changes.}
	\label{graph:tableaushape}
\end{figure}

The complexity of the \texttt{meld} algorithm for an input consisting
of $T$ terms each of which has an associated tableau $Y$ is of the
order $O(T N(Y))$ as the algorithm computes the projection once per
term.  On top of this is a linear decomposition problem which involves
inverting a $K \times K$ matrix where $K \leq T$ and checking a
solution against each term in the projection, but as the constant
factors for this part of the problem are far smaller we neglect this
from the analysis. In fig.~\ref{graph:n_terms} the linear dependence
on the number of terms is illustrated.

As with complexity analysis on other canonicalisation routines (see
e.g.~\cite{mans3},~\cite{Niehoff:2017mbk}) we look at the performance
as the number of indices in the input object increases. As there are
many different possible tableaux for each number of indices $n$ we
will discuss this for various tableaux shapes. The worst case
behaviour arises for fully (anti)symmetric tensors where the symmetry
is represented by a single row or column $S_n$ and has complexity
$N(S_n) = n!$ (although in practice there are many situations where
\texttt{meld} can optimise for this situation by simply sorting the
indices).

On the other hand a box tableau $B_n$, such as the Riemann tensor, has
$r_k = c_k = \sqrt{n}$ and thus $N(B_n) = \sqrt{n}!^{2\sqrt{n}}$. This
difference that the tableau shape makes to the complexity is shown in
fig.~\ref{graph:tableaushape}. As one gets closer to fully symmetric
or fully anti-symmetric objects, the runtime increases, and without
the optimisation for those extreme cases as discussed in
section~\ref{s:optimisations}, the difference between the shortest and
longest runtime would increase to more than a factor~$200$.

\begin{figure}[t]
\includegraphics[width=\textwidth]{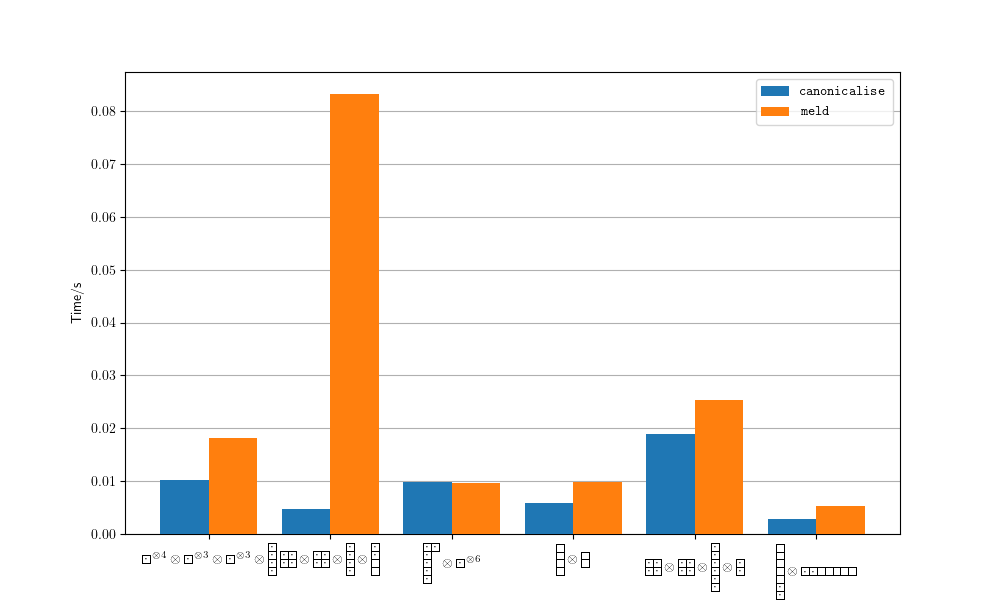}
\centering
\caption{Comparison of \texttt{meld} and \texttt{canonicalise} against
  various symmetry types. Boxes containing a mark represent contracted
  index slots.}
\label{graph:compare}
\end{figure}

We also examine the behaviour of the algorithm for various orders of
monomials as this is a common situation in fields such as perturbative
general relativity. Fig.~\ref{graph:polyorder} shows that scaling of
the algorithm against a line representing the scaling $O(a^n)$ for
monomial order $n$. The divergence from this at low orders is in
accordance with the constant overhead of the algorithm.

The fourth scaling we present is the behaviour as the fraction of
contracted index pairs varies. We briefly mentioned above the example
of~$S_{a b a b}$ where the symmetrisation produces only two terms, as
opposed to~$S_{a b c d}$ where 24~terms are produced. The effect this
has on the algorithm, presented in fig.~\ref{graph:dummy}, is harder to
quantify, although in general a higher fraction of contracted index
pairs results in a faster running time. Of the many factors which may
contribute to this, we draw attention to the following: fewer
allocations, more compact objects resulting in fewer cache misses, and
the reduction of the number of terms required to iterate over in the
calculation of linear dependence.

Obviously, this is not a behaviour which scales well for problems with
very large number of indices, but the purpose is to provide a
completely general solution for as many problems as possible rather
than optimising for niche problems and reducing the overall
generality.

Let us finally also present some practical considerations which result
from the above discussion by comparing the runtime performance on
expressions involving only mono-term symmetries, that is, for
situations in which both \texttt{meld} and Cadabra's
\texttt{canonicalise} algorithm (which uses
xPerm~\cite{DBLP:journals/corr/abs-0803-0862} under the hood) can be
used.  This of course necessarily limits attention to cases for which
\texttt{meld} was not primarly designed.  The \texttt{meld} algorithm,
while more brute-force, has the benefit of being relatively simple and
thus cache friendly, and its implementation is such that there is no
need to deal with dummy index permutations. The \texttt{canonicalise}
algorithm is more mathematically elegant, but because of that also
more complex, and at least in its current implementation needs to
consider a double-coset problem because dummy index relabelling needs
to be taken into account. The results are shown in
fig.~\ref{graph:compare} and show that in almost all scenarios the
\texttt{canonicalise} algorithm is a more efficient approach. We have
already mentioned this, but we wish to draw attention to the magnitude
of the difference.  In most cases the more complex
\texttt{canonicalise} algorithm outperforms \texttt{meld} by a factor
of less than two, which for everyday calculations may well be a minor
expense to be paid. Even in the cases when \texttt{meld} scales very
badly, the runtime of \texttt{meld} is still less than one-tenth of a
second, and in more catastrophic cases (which are omitted from the
graph to keep the other cases in scale) the actual runtime may only be
a few seconds.

Therefore, while there are certainly occasions in which the use of
\texttt{meld} is impractical, we find that as a general tool for
detecting symmetries it is often efficient enough for real-world
situations.

\section{Extensions}
\label{s:extensions}

The \texttt{meld} algorithm is, as previously mentioned, an algorithm
designed to be a general routine which can detect many different
symmetries without the user having to manually identify and apply
individual relations. The ability to detect multi-term relations
covers a large number of identities but by no means all
possibilities. Various extensions which could be considered are
presented here.

\subsection{Reduction of traces}

One such example which has been implemented is the cyclicity of the trace\,,
\begin{equation}
	\Tr(ABC) = \Tr(BCA) = \Tr(CAB)\,,
\end{equation}
where $A$, $B$ and $C$ do not necessarily commute. In the
\texttt{meld} implementation, each term in the trace is split into
commuting terms and non-commuting terms. The non-commuting terms are
``cycled'' until either they match against another term in the sum
using a dummy-agnostic comparison, or are cycled back to their initial
arrangement.

\subsection{Dimension dependent identities}

Another group of symmetries which fit into the objective of
\texttt{meld}, but are not currently implemented, are the
dimension-dependent identities which result from anti-symmetrisation of
an object with $n$ indices in $d < n$ dimensions,
\begin{equation}
	F_{[\alpha_1 \alpha_2 \dots \alpha_n]} = 0\,.
\end{equation}
This is a result of the limited number of ways of labelling the $n$ indices;  easily seen with
$n=2, d=1$ as the only way of labelling $F_{\alpha \beta} - F_{\beta \alpha}$ is $F_{x x} - F_{x x} = 0$.
From this, many more identities can be derived~\cite{dianyan} such as 
\begin{equation}
	R^a_{~b}R^{b c d e}R_{a c d e} = \frac{1}{4}RR^{a b c d}R_{a b c d} + 2R^{a c}R^{b d}R_{a b c d} 
	+ 2R^a_{~b}R^b_{~c}R^c_{~a} - 2RR^a_{~b}R^b_{~a} + \frac{1}{4}R^3\,,
\end{equation}
which can be calculated by expanding
\begin{equation}
	R^{ab}_{~~[ab}R^{cd}_{~~cd}R^e_{~e]} = 0\,.
\end{equation}
There exist general methods for generating such
identities~\cite{Edgar_2002} which are tedious calculations to perform
by hand and are excellent candidates for the types of relations a
routine like \texttt{meld} should detect. An extension of the
\texttt{meld} algorithm which deals with these identities will be
reported on elsewhere.

\subsection{Multi-term side relations}

As well as simplifying expressions based on the index symmetries of
tensors, another possibility for combining terms is by defining sets
of side relations and using these to reduce the size of the set of
basis terms. A motivating example for this is the commutator of
covariant derivatives in the absence of torsion,
\begin{equation}
	[\nabla_\mu, \nabla_\nu]V^\rho = R^\rho_{~\sigma\mu\nu}V^\sigma\,.
\end{equation}
In particular we might want to make use of
\begin{equation}
	\nabla_\mu \nabla_\nu R^\rho_{~\alpha\beta\gamma} = \nabla_\nu \nabla_\mu R^\rho_{~\alpha\beta\gamma} + 
	                                                    R^\rho_{~\sigma\mu\nu}R^\sigma_{~\alpha\beta\gamma}\,.
\end{equation}
Given an arbitrary expression containing second order derivatives of
Riemann tensors, it is a reasonable goal to use a basis of only two of
the terms in this expression. By expressing the relation in a form
which makes the side-relations manifest,
\begin{equation}
	\begin{aligned}
	\nabla_\mu \nabla_\nu R^\rho_{~\alpha\beta\gamma} &\equiv \frac{1}{2}(\nabla_\mu \nabla_\nu R^\rho_{~\alpha\beta\gamma}
                                                            +  \nabla_\nu \nabla_\mu R^\rho_{~\alpha\beta\gamma}  
	                                                        + R^\rho_{~\sigma\mu\nu}R^\sigma_{~\alpha\beta\gamma}	) \,,\\
	\nabla_\nu \nabla_\mu R^\rho_{~\alpha\beta\gamma} &\equiv \frac{1}{2}(\nabla_\mu \nabla_\nu R^\rho_{~\alpha\beta\gamma}
                                                            +  \nabla_\nu \nabla_\mu R^\rho_{~\alpha\beta\gamma}  
	                                                        - R^\rho_{~\sigma\mu\nu}R^\sigma_{~\alpha\beta\gamma}	) \,,\\
	R^\rho_{~\sigma\mu\nu}R^\sigma_{~\alpha\beta\gamma} &\equiv \frac{1}{2}(-\nabla_\mu \nabla_\nu R^\rho_{~\alpha\beta\gamma}
                                                            +  \nabla_\nu \nabla_\mu R^\rho_{~\alpha\beta\gamma}  
	                                                        + R^\rho_{~\sigma\mu\nu}R^\sigma_{~\alpha\beta\gamma}	) \,,\\
	\end{aligned}
\end{equation}
the similarity with the \texttt{meld} routine surfaces: each term in a sum is sequentially ``projected'' using the 
side-relations and an attempt made to write it as a linear combination of the previously projected terms. 

\section{Discussion and conclusion}

We have argued that canonicalisation is not necessarily the most
useful algorithm from the perspective of a computer algebra
\emph{user}, and will generally lead to expressions which are
unnecessarily large.  Instead of converting terms in an expression to
canonical form, it may be more useful to meld them together (in the
sense discussed in the main text), so that the form of the input
expression is preserved as much as possible. We have shown how this
can be made to work explitly for expressions involving tensors which
exhibit multi-term symmetries, and have briefly commented on
extensions to other related situations.

Our implementation uses a memory-efficient and dummy-name agnostic
storage method, and shows that many practical computations can be
handled in reasonable time. It is, however, not necessarily optimised
for speed yet, and could easily be improved by combining it with a
mono-term canonicaliser. Our focus has been on general applicability,
simplicity of the algorithm, and easy-of-use in practical
computations, for which the performance is certainly often good
enough, as the examples in section~\ref{sec:examples} show.

\newpage
\bibliographystyle{kasper}
\bibliography{meld}

\end{document}